\documentclass{article}

\usepackage{epsfig}
\usepackage{amssymb}
\usepackage{latexsym}
\usepackage{amsmath}
\usepackage{amsthm}

\newtheorem{theorem}{Theorem}
\newtheorem{proposition}{Proposition}
\newtheorem{definition}{Definition}

\date{}

\begin{document}

\title{Inertial forces and photon surfaces in arbitrary spacetimes}
\author{Thomas Foertsch\thanks{Mathematisches Institut, Universit\"at Z\"urich-Irchel, 
8057 Z\"urich, Switzerland. \newline 
Email: foertsch@math.unizh.ch} \hspace{0.3cm}
Wolfgang Hasse\thanks{Institut f\"ur Theoretische Physik, TU Berlin, Sekr. PN 7-1,
10623 Berlin, Germany, and  Wilhelm-Foerster-Sternwarte Berlin, 12169 Berlin, Germany.
Email: astrometrie@gmx.de} \hspace{0.3cm}
Volker Perlick\thanks{Institut f\"ur Theoretische Physik, Universit\"at zu K\"oln,
50923 K\"oln, Germany. Permanent address: Institut f\"ur Theoretische Physik, TU Berlin, 
Sekr. PN 7-1, 10623 Berlin, Germany. Email: vper0433@itp.physik.tu-berlin.de}}

\maketitle

\begin{abstract}
\noindent
Given, in an arbitrary spacetime $(M,g)$, a 2-dimensional timelike submanifold $\Sigma$
and an observer field $n$ on $\Sigma$, we assign gravitational, centrifugal, Coriolis and Euler 
forces to every particle worldline $\lambda$ in $\Sigma$ with respect to $n$. We prove that 
centrifugal and Coriolis forces vanish, for all $\lambda$ in $\Sigma$ with respect to 
any $n$, if and only if $\Sigma$ is a photon 2-surface, i.e., generated by two families 
of lightlike geodesics. We further demonstrate that a photon 2-surface can be 
characterized in terms of gyroscope transport and we give several mathematical 
criteria for the existence of photon 2-surfaces. Finally, examples of photon 2-surfaces 
in conformally flat spacetimes, in Schwarzschild and Reissner-Nordstr\"om spacetimes,
and in G\"odel spacetime are worked out.
\\[0.2cm]
\end{abstract}

\vspace{0.5cm}

%%%%%%%%%%%%%%%%%%%%%%%%%%%%%%%%%%%%%%%%%%%%%%%%%%%%%%%%%%%%%%%%%%%%%%%%%%%%%%%%%%%%%%%%%%%%%
%%%%%%%%%%%%%%%%%%%%%%%%%%%%%%%%%%%%%%%%%%%%%%%%%%%%%%%%%%%%%%%%%%%%%%%%%%%%%%%%%%%%%%%%%%%%%

\section{Introduction}
\label{sec-intro}

In this paper we discuss the question of how to define inertial forces in an arbitrary 
general-relativistic spacetime, i.e., in a Lorentzian manifold $(M,g)$.  Given a particle's 
worldline $\lambda$ in $M$, the 4-acceleration of $\lambda$ is unambiguously
defined.  The negative 4-acceleration of $\lambda$ gives the acceleration of a freely 
falling particle relative to $\lambda$.   After multiplication with the freely 
falling particle's mass we get the total ``apparent force'' that acts, in the view of 
$\lambda$, upon the freely falling particle.  By the question of how to define inertial 
forces we mean the question of how to decompose this total ``apparent force'' into 
gravitational, centrifugal, Coriolis and Euler force. This decomposition requires some 
additional information, i.e., it cannot be done in an absolute way. (The statement that 
the gravitational force has no absolute meaning is one version of the equivalence principle.) 
In a static or stationary spacetime, this additional information is provided by the 
timelike Killing vector field which, after normalization, can be interpreted as the 
4-velocity field of a distinguished family of observers. With respect to this distinguished 
observer field, inertial forces can be uniquely assigned to each particle's worldline, as 
was discussed by Abramowicz and collaborators in various articles, see, e.g. Abramowicz, 
Carter and Lasota \cite{acl}, Abramowicz and Prasanna \cite{ap} and Abramowicz \cite{a}. 
They also provided a geometric interpretation of inertial forces in terms of the so-called 
reference-3-geometry that was obtained by conformally  adjusting the quotient-space metric. 
For a discussion of these ideas we also refer to a forthcoming book by Abramowicz 
and Sonego \cite{as}.

The results in static and stationary spacetimes led Abramowicz and others 
to seek for particular observer fields in general spacetimes that allowed 
to define similar reference-3-geometries and thus to generalize the notion 
of inertial forces to general spacetimes. 
However, their suggested generalizations (see Abramowicz, Nurowski and 
Wex \cite{anw1}) have been partially criticized by several authors  
(see for example de Felice \cite{fe}, Sonego and Massar \cite{sm} and 
Bini, Carini and Jantzen \cite{bcaj2}). Some of the critical comments
refer to the fact that the inertial forces, as defined in \cite{anw1},
have unwanted features, others to the fact that they are in a certain
sense ambiguous.  Thus, it seems fair to say that the appropriate definition 
of inertial forces in arbitrary spacetimes is still a matter of debate.

In this paper we turn away from the idea that in arbitrary spacetimes
inertial forces should be defined with respect to an observer
field with particular properties. We rather define inertial forces
with respect to arbitrary timelike surfaces $\Sigma$ 
and arbitrary observer fields given on $\Sigma$. More precisely, 
we choose, in an arbitrary spacetime $(M,g)$, a timelike surface 
$\Sigma$ and a vector field $n$ on $\Sigma$ that is normalized to 
$g(n,n)=-1$. We demonstrate that then gravitational, centrifugal, Coriolis 
and Euler forces can be unambiguously assigned to any particle's
worldline $\lambda$ contained in $\Sigma$. These definitions will
be given in Section \ref{sec-inertial} below. 

For the sake of illustration, one may interpret $\Sigma$ as the 
worldsheet of a roller-coaster track and $n$ as the 4-velocity field 
of observers distributed along the track. Then the worldlines $\lambda$ 
contained in $\Sigma$ can be associated with the motions allowed for cars 
that are bound to the roller-coaster track but are free to move arbitrarily
along the track. This roller-coaster interpretation, which was used  
e.g. by Abramowicz in \cite{a}, is quite helpful for associating inertial 
forces with physical intuition. However, we emphasize that the roller-coaster 
interpretation is only a didactical means. For our mathematical results to be 
true it is not necessary to associate $\Sigma$ with a material object.

The idea of defining the centrifugal force with respect to a  
prescribed spatial track can be traced back to Huygens, whereas the
definition usually given in mechanics text-books is due to Newton, 
see Abramowicz \cite{a}, p.735/736. Our definitions of inertial 
forces in Section \ref{sec-inertial} can be viewed as a literal adaptation 
of Huygens' definition of centrifugal force to general relativity. We
share Abramowicz's view that, in the context of general relativity,
Newton's definition is not appropriate.
 
If a timelike vector field $K$ has been chosen on $(M,g)$, every timelike 
curve $\lambda$ that is nowhere tangent to $K$ naturally defines a timelike 
surface $\Sigma$, namely the set of all points that can be connected to 
$\lambda$ by an integral curve of $K$. Restricting to $\Sigma$ and normalizing 
$K$ gives a vector field $n$ on $\Sigma$. In other words, once a timelike 
vector field $K$ has been chosen on $(M,g)$ we have a distinguished choice 
for $(\Sigma ,n)$ and, thus, unique inertial forces for every worldline 
$\lambda$ that is nowhere tangent to $K$. (For a $\lambda$ that is 
tangent to $K$, at some points or everywhere, we may choose any timelike 
surface $\Sigma$ that contains $\lambda$ and is invariant under the flow 
of $K$; the inertial forces defined along $\lambda$ turn out to be 
the same for all these choices.) The crucial point is that $K$   
need not satisfy any additional condition. In the special case that 
$K$ is a Killing vector field our definition of inertial forces reduces 
to the generally accepted one in static or stationary spacetimes. 

Having defined inertial forces with respect to a timelike 
surface $\Sigma$, it is an interesting problem to determine those
$\Sigma$ for which centrifugal and Coriolis forces vanish.
We do this in Section \ref{sec-theorems} by proving the

\begin{theorem} \label{theo-main}
Let $(M,g)$ be an arbitrary spacetime and $\Sigma$ a $2$-dimensional 
timelike submanifold of $M$. Then the following properties are
mutually equivalent:
\begin{description}
\item[(i)] For every particle worldline $\lambda$ contained in 
$\Sigma$ the centrifugal and Coriolis forces with respect
to any vector field $n$ on $\Sigma$ with $g(n,n)=-1$ vanish.   
\item[(ii)] $\Sigma$ is generated by two congruences of lightlike 
geodesics.
\item[(iii)]
If $n$ and $\tau$ are any two vector fields on $\Sigma$ 
with $g(\tau , \tau ) = - g(n,n) = 1$ and $g(n, \tau) =0$, the
vector field $\tau$ is Fermi-Walker-parallel along $n$, i.e., $\nabla
_n \tau$ is a multiple of $n$.
\end{description}
\end{theorem}

If we interpret $\Sigma$ as the worldsheet of a track, Property
(ii) means that the track appears straight to the eye of an
observer who looks (``backward'' or ``forward'') along the track.
As, according to general relativity, a gyroscope always remains 
Fermi-Walker-parallel to itself (see, e.g. Misner, Thorne and 
Wheeler \cite{mtw}, Sect. 40.7), Property (iii) means that when 
transported along the track a gyroscope whose axis is initially 
parallel to the track always stays parallel to the track.

Note that, although the centrifugal and Coriolis forces depend on the choice of $n$, 
Theorem \ref{theo-main} demonstrates that the condition of vanishing 
centrifugal and Coriolis forces is independent of $n$, i.e., it is a 
geometric property of $\Sigma$. Moreover, Theorem \ref{theo-main} 
implies that properties (i) and (iii) are conformally invariant  
because property (ii) obviously is. The conformal invariance of
property (iii) reflects the known fact that the notion of an 
``inertial compass'' is conformally invariant.

A trivial example for a timelike surface $\Sigma$ that obviously 
satisfies Property (ii) and, thus, Properties (i) and (iii) of 
Theorem \ref{theo-main} is a timelike plane in Minkowski spacetime.  
The best known non-trivial example is the surface $r = 3M$, $\theta = \pi /2$ 
in the Schwarzschild spacetime. If we interpret this surface as the worldsheet 
of a track, someone moving along this track feels neither centrifugal nor Coriolis 
forces, although the track is circular; at the same time, Properties (ii) and (iii) 	
demonstrate that the track appears straight to the eye and turns out to be straight 
if deviation from straightness is measured by transporting a gyroscope.
 
In generalizing this well-known Schwarzschild example, Abramowicz \cite{a} 
has considered timelike surfaces that are invariant under the flow of the 
Killing vector fields $\partial _t$ and $\partial _{\varphi}$ in a static
axisymmetric spacetime. He has shown that in this case Property (i) of
Theorem \ref{theo-main} implies Properties (ii) and (iii). Our result 
generalizes this observation to arbitrary timelike surfaces in arbitrary 
spacetimes.  

Theorem \ref{theo-main} gives three equivalent characterizations of
surfaces but it does not say anything about the existence of such
surfaces in arbitrary spacetimes. We call a 2-dimensional timelike 
submanifold $\Sigma$ that satisfies Property (ii) and, thus, 
Properties (i) and (iii) of Theorem \ref{theo-main} a {\em timelike 
photon $2$-surface}. A similar notion was used by Claudel, 
Virbhadra and Ellis \cite{cve}. In arbitrary spacetimes, the existence 
of timelike photon 2-surfaces is not guaranteed. Some existence 
criteria and construction methods are presented in Section 
\ref{sec-photonsurface}. Finally in Section \ref{sec-examples} 
we give various examples for timelike photon 2-surfaces.

%%%%%%%%%%%%%%%%%%%%%%%%%%%%%%%%%%%%%%%%%%%%%%%%%%%%%%%%%%%%%%%%%%%%%%%%%%%%%%%%%%%%%%%%%%%%%
%%%%%%%%%%%%%%%%%%%%%%%%%%%%%%%%%%%%%%%%%%%%%%%%%%%%%%%%%%%%%%%%%%%%%%%%%%%%%%%%%%%%%%%%%%%%%

\section{The definition of inertial forces}
\label{sec-inertial}

In this section we define the inertial forces in an arbitrary spacetime. Here,
by a spacetime we mean any finite-dimensional Lorentzian manifold $(M,g)$. For physical 
reasons, $(M,g)$ should be 4-dimensional, time-orientable, and connected; mathematically,
however, we shall not need these particular assumptions. Our results are non-trivial 
only if $M$ is at least 3-dimensional.

Throughout this paper, our conventions 
are as follows. We use physical units making the velocity of light equal to one and 
our sign convention for the Lorentzian signature is $(+,\dots ,+,-)$. The Levi-Civita
derivative of $g$ is denoted by $\nabla$. All (sub)manifolds and maps are tacitly assumed 
to be sufficiently often differentiable such that the written derivatives exist.
By a submanifold we mean what is more fully called an immersed submanifold, as opposed
to an imbedded submanifold, i.e., we allow for self-intersections. This is irrelevant 
for the definition of inertial forces, which is purely local, and for the three conditions 
in Theorem \ref{theo-main}, whose equivalence can be verified for arbitrarily small pieces 
of $\Sigma$; however, it is important to note that our construction methods for photon 
surfaces may yield surfaces with self-intersections. For a submanifold $\Sigma$, we
distinguish between vector fields {\em on\/} $\Sigma$ and vector fields {\em along\/} $\Sigma$.
A vector field on $\Sigma$ assigns to each point $p \in \Sigma$ a vector in the tangent space 
$T_p \Sigma$ whereas a vector field along $\Sigma$ assigns to $p \in \Sigma$ a vector in the 
tangent space $T_{i(p)} M$ where $i$ denotes the immersion $\Sigma \rightarrow M$. (An analogous 
terminology applies to tensor fields of higher rank.) To make the distinction easily seen we use 
lower-case letters for vector fields on a submanifold of $M$ and upper-case letters for vector 
fields that take values in the full tangent space of $M$.     

What we need in order to define the inertial forces is a pair $(\Sigma ,n)$, 
where $\Sigma$ is a 2-dimensional timelike submanifold of $M$, and $n$ is a 
vector field on $\Sigma $ with $g(n,n)=-1$. Given any timelike curve $\lambda 
: I_{\lambda}\longrightarrow \Sigma$ whose tangent vector field $u= {\dot{\lambda}}$ 
satisfies the usual 4-velocity normalization condition $g(u,u)=-1$, it is our goal 
to assign inertial accelerations to $\lambda$ with respect to $(\Sigma ,n)$. The 
inertial forces acting upon a particle are then   
given by multiplying the inertial accelerations with the particle's mass.

With $(\Sigma ,n)$ given, the conditions
\begin{equation}\label{eqn-deftau}
g(n, \tau ) \, = \, 0 \qquad \text{and} \qquad g(\tau , \tau ) \, = \, 1
\end{equation}
define a vector field $\tau$ on $\Sigma$ uniquely up to sign. After fixing 
$\tau$ by choosing one of the two signs, we can decompose the 4-velocity $u$ 
of each worldline $\lambda$ in the form
\begin{equation}\label{eqn-defv}
u \; = \; \gamma \, (n \; + \; v \, \tau ) \, , \quad
\gamma \, = \, \frac{1}{\sqrt{1-v^2}} \, .
\end{equation}
This equation defines the spatial velocity $v$ of $\lambda$ with respect to $n\,$.
Mathematically, $v$ is a map that assigns to the curve parameter a number 
between $-1$ and 1. From (\ref{eqn-defv}) we can calculate the 4-acceleration 
$\nabla _u u$ of $\lambda$,
\begin{equation}\label{eqn-defa}
\begin{split}
{\nabla}_uu \; = \; {\nabla}_nn \; + \; {\gamma}^2 v \Big( {\nabla}_n\tau \; + 
\; {\nabla}_{\tau}n\Big) \; + \;
\\ 
{\gamma}^2v^2 \Big( {\nabla}_{\tau}\tau \; + \; {\nabla}_nn\Big) \; 
 + \; {\nabla}_u(\gamma v) \, \tau \; + \; ({\nabla}_u\gamma ) n  \; .
\end{split}
\end{equation}
As $\nabla _u u$ gives the acceleration of $\lambda$ relative to freely 
falling particles, relative to $\lambda$ a freely falling particle is 
subject to the ``inertial acceleration'' $a=-\nabla _u u$.  
The gravitational, centrifugal, Coriolis and Euler accelerations with
respect to the observer field $n$ are obtained by the following unique 
decomposition of $a$ (and multiplication with the freely falling particle's
mass gives the respective forces).
\begin{description}
\item[(a) ] We decompose $a$ into a component orthogonal to $n$ and  
a component parallel to $n$. By (\ref{eqn-defa}), the component orthogonal 
to $n$ is
\begin{equation}\label{eqn-aperp}
\begin{split}
a \, + \, g(a,n) \, n \;  = \;- \,  {\nabla}_nn \; - \; {\gamma}^2 v \Big( {\nabla}_n\tau \; + 
\; {\nabla}_{\tau}n \; + \; g({\nabla}_n\tau ,n) \, n \Big) 
\; - 
\\
 {\gamma}^2 v^2 \Big(  {\nabla}_{\tau}\tau \; + \;
{\nabla}_{n}n \; + \; g({\nabla}_{\tau}\tau ,n) \, n \Big) \; 
- \; \Big( {\nabla}_u(\gamma v)\Big) \, \tau \; .
\end{split}
\end{equation}
\item[(b) ] The {\em gravitational acceleration\/} $a_{\mathrm{gra}}$ is by definition the 
part of $a \, + \, g(a,n) \, n$ that is independent of $v$, i.e., the term on the right-hand 
side of (\ref{eqn-aperp}) that survives if $v$ is identically zero,  
\begin{equation}\label{eqn-agra}
a_{\mathrm{gra}} \; = \; - \, {\nabla}_nn \; .
\end{equation}
\item[(c) ] The remaining part $a \, + \, g(a,n) \, n + {\nabla}_nn$ will be 
decomposed into the following pieces:  
\begin{description}
\item[(i)] The part orthogonal to ${\tau}$ and odd in $v$ is the 
{\em Coriolis acceleration}
\begin{equation}\label{eqn-acor}
a_{\mathrm{Cor}} \; = \; - \, {\gamma}^2 v 
\Big( {\nabla}_n\tau \; + \; {\nabla}_{\tau}n \; + 
\; g({\nabla}_n\tau ,n) \, n \; - \; g({\nabla}_{\tau}n,\tau ) 
\, \tau \Big) \, .
\end{equation}
\item[(ii)]  The part orthogonal to ${\tau}$ and even in $v$ is the 
{\em centrifugal acceleration}
\begin{equation}\label{eqn-acen}
a_{\mathrm{cen}} \; = \; - \, {\gamma}^2 v^2 
\Big(  {\nabla}_{\tau}\tau \; + \; {\nabla}_{n}n \; + 
\; g({\nabla}_{\tau}\tau ,n) \, n \; - 
\; g({\nabla}_nn,\tau ) \, \tau \Big) \, .
\end{equation}
\item[(iii)] The part parallel to $\tau$ is the {\em Euler acceleration} 
\begin{equation}\label{eqn-aeul}
a_{\mathrm{Eul}} \; = \; - \, 
\Big( {\nabla}_u(\gamma v) \; - \; {\gamma}^2 v \, g({\nabla}_{\tau}\tau ,n) 
\; + \; \gamma ^2 v^2 g({\nabla}_nn,\tau ) \Big) \, \tau \, . \quad
\end{equation}
\end{description}
\end{description}
Note that this decomposition ensures that two worldlines $\lambda _1$
and $\lambda _2$ with $\lambda _1 (0) = \lambda _2 (0) = p$ and
$v_1 (0)=-v_2(0)$ have the same centrifugal accelerations but
opposite Coriolis accelerations at $p$.

Whereas $a_{\mathrm{cen}}$ and $a_{\mathrm{Cor}}$ are always perpendicular
to $\Sigma$ and $a_{\mathrm{Eul}}$ is always parallel to $\Sigma$,  
the gravitational acceleration has, in general, a component perpendicular and a
component parallel to $\Sigma$, say $a_{\mathrm{gra}} \, = \, 
a _{\mathrm{gra}} ^{\perp} \, + \, a_ {\mathrm{gra}}^{\|} \,$. Thus, the components 
of $a$ perpendicular and parallel to $\Sigma$ are, respectively,
\begin{equation}\label{eqn-aop}
a ^{\perp}\, = \, a_{\mathrm{gra}}^{\perp} \, + \, 
a _{\mathrm{cen}} \, + \, a_{\mathrm{Cor}}  \; , \qquad 
a ^{\|}\, = \, a_{\mathrm{gra}}^{\|} \, + \, a_{\mathrm{Eul}} \, - \, g(a,n) \, n \; .
\end{equation}
Clearly, $a ^{\perp}$ and $a ^{\|}$ are invariant with respect to changing the
observer field $n$. As $a \, = \, - \, \nabla _u u$ is orthogonal to $u$, both 
$a ^{\perp}$ and $a ^{\|}$ are spatial vectors in the local rest system of the 
particle $\lambda$. If we interpret $\Sigma$ as the worldsheet of a track, 
$a ^{\perp}$ gives the inertial acceleration perpendicular to the track 
whereas $a ^{\|}$ gives the inertial acceleration parallel to the track. The 
latter involves a derivative of the velocity  that enters into the Euler 
force; hence, it depends on whether the particle accelerates or decelerates in 
the direction of the track. As a consequence, there is no general relation 
between  $a ^{\|}$ and the geometry of $\Sigma$. For our Theorem \ref{theo-main}, 
and other results that relate the inertial forces to the geometry of $\Sigma$, 
it is only $a ^{\perp}$ that matters. Note that, since $a_{\mathrm{cen}}$ and 
$a_{\mathrm{Cor}}$ are spatial vectors in the local rest system of the 
particle $\lambda$, our Theorem \ref{theo-main} is directly related to 
inertial accelerations which can be measured in this system.

Recall that, with $(\Sigma, n)$ given, $\tau$ was determined by 
(\ref{eqn-deftau}) uniquely up to sign. If we change $\tau$ into $- \tau$, 
(\ref{eqn-defv}) shows that $v$ changes into $-v$. With this observation 
it is obvious from (\ref{eqn-agra}), (\ref{eqn-acor}), (\ref{eqn-acen}) 
and (\ref{eqn-aeul}) that $a_{\mathrm{gra}}$, $a_{\mathrm{Cor}}$, 
$a_{\mathrm{cen}}$ and $a_{\mathrm{Eul}}$ are indeed uniquely determined 
by $(\Sigma , n)$ for each worldline $\lambda$ in $\Sigma$. This is an 
important difference in comparison to the work of Abramowicz, Nurowski and 
Wex \cite{anw1} where $\tau$ was uniquely determined only along $\lambda$ 
and an ambiguity in the definition of inertial forces arose from the 
necessity of extending $\tau$ off $\lambda \,$. Another important difference 
is in the fact that the centrifugal force defined in \cite{anw1} 
need not be perpendicular to the plane spanned by the particle's 
4-velocity and the observer field.

%%%%%%%%%%%%%%%%%%%%%%%%%%%%%%%%%%%%%%%%%%%%%%%%%%%%%%%%%%%%%%%%%%%%%%%%%%%%%%%%%%%%%%%%%%%%%
%%%%%%%%%%%%%%%%%%%%%%%%%%%%%%%%%%%%%%%%%%%%%%%%%%%%%%%%%%%%%%%%%%%%%%%%%%%%%%%%%%%%%%%%%%%%%

\section{A characterization of worldsheets for which centrifugal and Coriolis
forces vanish}
\label{sec-theorems}

It is the main purpose of this section to prove Theorem \ref{theo-main} which 
was announced in the Introduction. To that end we have to introduce some 
notation. Given a 2-dimensional timelike submanifold $\Sigma$ in our Lorentzian
manifold $(M,g)$, we decompose at each point $p \in \Sigma$ the tangent space 
$T_p M$ into $T_p \Sigma$ and its orthocomplement $(T_p \Sigma)^{\perp}$. 
With $p$ running over $\Sigma$, the orthogonal projections
\begin{equation}\label{eqn-Pp}
P_p^{\perp}: T_pM \, \longrightarrow \, (T_p\Sigma )^{\perp} \, ,
\end{equation}
define a tensor field $P^{\perp}$ along $\Sigma$ that maps vector fields 
along $\Sigma$ to vector fields along $\Sigma$. After choosing a vector 
field $n$ with $g(n,n)=-1$ on $\Sigma$ and denoting by $\tau$ the
vector field on $\Sigma$ that is determined by (\ref{eqn-deftau})
uniquely up to sign, $P^{\perp}$ takes the form
\begin{equation}\label{eqn-Pntau}
P^{\perp}(Y) = Y - g(\tau,Y) \, \tau + g(n,Y) \, n \, ,
\end{equation}
where $Y$ is an arbitrary vector field along $\Sigma$. 
The Coriolis acceleration (\ref{eqn-acor}) and the centrifugal acceleration 
(\ref{eqn-acen}) can be rewritten, with the help of $P^{\perp}$, in the 
following way.
\begin{equation}\label{eqn-corcen}
a_{\mathrm{Cor}} \; = \; - \, {\gamma}^2 v \, P^{\perp} 
\Big( {\nabla}_n\tau \; + \; {\nabla}_{\tau}n\Big) \hspace{0.5cm}
\mbox{and} \hspace{0.5cm}
a_{\mathrm{cen}} \; = \; - \, {\gamma}^2v^2 \, P^{\perp} \Big( {\nabla}_nn 
\; + \; {\nabla}_{\tau}\tau \Big) .
\end{equation}
Since $n$ and $\tau$ are orthonormal, the equation 
\begin{equation}\label{eqn-defl}
l^{\pm} \, = \, n \, \pm \, \tau
\end{equation}
defines two lightlike vector fields $l^+$ and $l^-$ on $\Sigma$.
At each point $p$ of $\Sigma$, the vectors $l_p^+$ and $l_p^-$ span 
the two different lightlike lines tangent to $\Sigma$. Note that 
$l^+$ and $l^-$ satisfy
\begin{equation}\label{eqn-lnorm}
g(l^+, l^-) \, = \, - \, 2 \, .
\end{equation}
If $n$ has been chosen, $\tau$ is unique up to sign, so $l^+$ and
$l^-$ are unique up to interchanging.
If we replace $n$ by some other vector field $\tilde {n}$ on $\Sigma$ 
with $g({\tilde{n}},{\tilde{n}}) = -1$ (and $\tau$ correspondingly
by $\tilde{\tau}$), the vector fields $l^+$ and $l^-$ transform 
according to
\begin{equation}\label{eqn-transl}
l^+ \longmapsto A \, l^+ \quad {\mathrm{and}} \quad
l^- \longmapsto A^{-1} \, l^- \, ,
\end{equation}
where $A$ is a nowhere vanishing scalar function on $\Sigma$.
With the help of (\ref{eqn-defl}), the expression 
(\ref{eqn-Pntau}) for the projection tensor field $P^{\perp}$
can be rewritten as
\begin{equation}\label{eqn-Pl}
P^{\perp} (Y) \, = \, Y \, + \, \tfrac{1}{2} \, g(l^-,Y) \, l^+ 
\, + \, \tfrac{1}{2} \, g(l^+,Y) \, l^- \, .
\end{equation}
As $l^+$ and $l^-$ are lightlike, (\ref{eqn-Pl})
immediately implies
\begin{equation}\label{eqn-dell}
P^{\perp} \big( \nabla _{l^{\pm}} l^{\pm} \big) \, = \, 
\nabla _{l^{\pm}} l^{\pm} \, + \, \tfrac{1}{2} \, 
g \big( l^{\mp} , \nabla _{l^{\pm}} l^{\pm} \big) \, l_{\pm} \, .
\end{equation}
After these preparations it is now easy to prove the following
proposition which will then be used to prove Theorem
\ref{theo-main}.

\begin{proposition} \label{prop-general}
Let $(M,g)$ be an arbitrary spacetime and $\Sigma$ a $2$-dimensional 
timelike submanifold of $M$. For any choice of a vector field
$n$ on $\Sigma$ with $g(n,n)=-1$, the equation 
\begin{equation}\label{eqn-cencrit} 
a_{\mathrm{cen}} \; = \; \mp \, v \, a_{\mathrm{Cor}} 
\end{equation}
holds for all particle worldlines $\lambda$ in $\Sigma$ with the upper sign 
$($or with the lower sign, resp.$)$ if and only if the integral curves of 
$l^+$ $($or of $l^-$, resp.$)$ are geodesics.    
\end{proposition}
\begin{proof}
From (\ref{eqn-corcen}) and (\ref{eqn-defl}) we read
\begin{equation}\label{eqn-al}
a_{\mathrm{cen}} \, \pm \, v \, a_{\mathrm{Cor}} \, = \, - \, 
\gamma ^2 \, v^2 \, P^{\perp} \big( \nabla _{l^{\pm}} l^{\pm} \big) \, .
\end{equation}
If the integral curves of $l^{\pm}$ are geodesics, $\nabla _{l^{\pm}}
l^{\pm}$ is a multiple of $l^{\pm}$ and, thus, tangent to $\Sigma$,
so the right-hand side of (\ref{eqn-al}) vanishes. Hence, the 
left-hand side has to vanish. Conversely, assume that the left-hand
side of (\ref{eqn-al}) vanishes for all particle worldlines in $\Sigma$.
Then the right-hand side of (\ref{eqn-al}) has to vanish for all allowed
$v$, i.e., $P^{\perp} \big( \nabla _{l^{\pm}} l^{\pm} \big)$ 
has to vanish. By (\ref{eqn-dell}), this implies that $\nabla _{l^{\pm}} 
l^{\pm}$ is a multiple of $l^{\pm}$.
\end{proof}

As the proof of Proposition \ref{prop-general} is purely algebraic,
we have actually proven that (\ref{eqn-cencrit}) holds at a 
particular point $p$ of $\Sigma$, for any choice of $n$ and for all
particle worldlines passing through $p$, if and only if 
$\nabla _{l^{\pm}} l^{\pm}$ is a multiple of $l^{\pm}$ at this 
particular point $p$.

With Proposition \ref{prop-general} at hand, we can now prove
our main theorem.

\vspace{0.2cm}
\noindent
{\em Proof of Theorem\/} \ref{theo-main}.
Proposition \ref{prop-general} implies that the integral curves
of both $l^+$ and $l^-$ are geodesics if and only if (\ref{eqn-cencrit})
holds with both signs for all particle worldlines $\lambda$ in $\Sigma$. The 
latter condition is, of course, true if and only if $a_{\mathrm{cen}}$ and  
$a_{\mathrm{Cor}}$ vanish for all particle worldlines $\lambda$ in $\Sigma$. 
This proves the equivalence of (i) and (ii) in Theorem \ref{theo-main}. -- 
To prove the equivalence of (ii) and (iii), we start from the equation
\begin{equation}\label{eqn-delntau}
4 \, \nabla _n \tau \, = \, \nabla _{l^+} l^+ \, - \, 
\nabla _{l^-} l^- \, - \, [l^+ , l^-] \, ,
\end{equation}
which follows directly from (\ref{eqn-defl}). As with $l^+$ and
$l^-$ also their Lie bracket must be tangent to $\Sigma$, this
implies
\begin{equation}\label{eqn-Pdelntau}
4 \, P^{\perp} \big( \nabla _n \tau \big) \, = \, 
P^{\perp} \big( \nabla _{l^+} l^+ \, - \, 
\nabla _{l^-} l^- \big) \, .
\end{equation}
When changing to new orthonormal vector fields ${\tilde{n}}$
and ${\tilde{\tau}}$, the vector fields $l^+$ and $l^-$ transform
according to (\ref{eqn-transl}), so
\begin{equation}\label{eqn-Pdelntautrans}
4 \, P^{\perp} \big( \nabla _{\tilde{n}} {\tilde{\tau}} \big) \, = \, 
A^2 \, P^{\perp} \big( \nabla _{l^+} l^+ \big) \, - \, 
A^{-2} \, P^{\perp} \big( \nabla _{l^-} l^- \big) \, .
\end{equation}
Owing to the normalization condition $g({\tilde{\tau}}, {\tilde{\tau}})
=1$, the vector field $\nabla _{\tilde{n}} {\tilde{\tau}}$ is 
automatically perpendicular to ${\tilde{\tau}}$; hence, this vector 
field is parallel to ${\tilde{n}}$ if and only if it is annihilated by $P^{\perp}$.
As a consequence, Property (iii) of Theorem \ref{theo-main} is
satisfied if and only if the right-hand side of (\ref{eqn-Pdelntautrans})
gives zero for all nowhere vanishing functions $A$ on $\Sigma$. By
(\ref{eqn-dell}), this is true if and only if the integral curves of
$l^+$ and $l^-$ are geodesics. 
\hfill $\Box$

\vspace{0.2cm}
In an arbitrary spacetime, it is easy to construct a timelike 
surface $\Sigma$ such that the integral curves of either 
$l^+$ or $l^-$ are geodesics: Just choose any lightlike and geodesic 
vector field $L$ on $M$ and a curve $\lambda$ which 
is nowhere orthogonal to $L$; then let $\Sigma$ be the 
set of all points in $M$ that can be connected to $\lambda$ by an
integral curve of $L$. By construction, $\Sigma$ is a 2-dimensional
immersed timelike submanifold of $M$ near $\lambda$ which is ruled
by a congruence of geodesics, namely by integral curves of $L$. 
The other lightlike congruence in $\Sigma$, however, will not be 
geodesic in general. In the next section we give criteria for the 
existence of timelike surfaces for which both $l^+$ and $l^-$ are
geodesic.

%%%%%%%%%%%%%%%%%%%%%%%%%%%%%%%%%%%%%%%%%%%%%%%%%%%%%%%%%%%%%%%%%%%%%%%%%%%
%%%%%%%%%%%%%%%%%%%%%%%%%%%%%%%%%%%%%%%%%%%%%%%%%%%%%%%%%%%%%%%%%%%%%%%%%%%
\section{The existence of photon surfaces}
\label{sec-photonsurface}
 
We begin with the following definition.

\begin{definition}\label{def-photonsurface}
A $k$-dimensional submanifold $\Sigma$ of a Lorentzian
manifold $(M,g)$ is called a {\em photon $k$-surface\/} if 
$\Sigma$ is nowhere spacelike and if for every $p \in \Sigma$ 
and every lightlike vector $w \in T_p \Sigma$ the geodesic 
$\mu : \; ] \, - \varepsilon , \varepsilon \, [ \; 
\longrightarrow M$ with ${\dot{\mu}} (0) =w$ is contained
in $\Sigma$, for some $\varepsilon > 0$.
\end{definition}

This definition slightly modifies the notion of ``photon
surfaces'' as it was extensively discussed by Claudel,
Virbhadra and Ellis \cite{cve}. The modification is in
the fact that Claudel, Virbhadra and Ellis consider only
hypersurfaces, i.e., submanifolds of codimension one.

It is obvious from Definition \ref{def-photonsurface} that
a 2-dimensional timelike submanifold is a photon 2-surface
if and only if it is generated by two families of lightlike
geodesics. Thus, Theorem \ref{theo-main} provides two additional
characterizations of timelike photon 2-surfaces, as an alternative 
to the one by lightlike geodesics . 

In this section we want to present some general results that guarantee
the existence of timelike photon 2-surfaces in some classes of spacetimes 
and tell us how to construct them. A quite obvious criterion is 
formulated in the following proposition.

\begin{proposition}\label{prop-frobenius}
A point $p$ in a Lorentzian manifold $(M,g)$ admits a neighborhood that
can be foliated into timelike photon $2$-surfaces if and only if on some
neighborhood of $p$ there are two linearly independent lightlike and 
geodesic vector fields $L^+$ and $L^-$ such that the Lie bracket $[L^+,L^-]$ 
is a linear combination of $L^+$ and $L^-$. In this case, the photon 
$2$-surfaces are the integral manifolds of the $2$-spaces spanned by 
$L^+$ and $L^-$.
\end{proposition}
\begin{proof}
Two linearly independent lightlike vector fields $L^+$ and $L^-$ span
a timelike 2-space at each point. By the well-known Frobenius theorem 
(see e.g. Westenholz \cite{we}, Proposition 3.13), these 2-spaces
admit local integral manifolds if (and only if) the Lie bracket of 
$L^+$ and $L^-$ is a linear combination of $L^+$ and $L^-$. The
condition of $L^+$ and $L^-$ being geodesic makes sure that these
integral manifolds are timelike photon 2-surfaces. This proves the
`if'-part. To prove the `only if'-part one just has to verify that
the lightlike and geodesic vector fields $l^+$ and $l^-$, given on 
each leave of the foliation up to transformations (\ref{eqn-transl}),
can be chosen such that they make up two smooth vector fields $L^+$
and $L^-$ on some neighborhood of $p$. 
% see e.g. Schutz \cite{sc}, p. 81.
\end{proof}

In Subsections \ref{subsec-schwarzschild} and \ref{subsec-goedel} below
we use Proposition \ref{prop-frobenius} for constructing families of 
photon 2-surfaces in Schwarzschild, Reissner-Nordstr\"om and G\"odel spacetime.
In more complicated spacetimes, however, it is rather difficult to find out 
whether or not two vector fields $L^+$ and $L^-$ with the desired properties 
do exist. In some cases of interest it could be of help to use the Newman-Penrose 
formalism, in particular the modified version of this calculus, due to Geroch,
Held and Penrose \cite{ghp}, where a pair of real lightlike vectors is 
picked out at each point. 

As an alternative to Proposition \ref{prop-frobenius} we will now give 
a construction method for timelike photon 2-surfaces in Lorentzian manifolds which 
admit a conformal Killing vector field. To work this out, we first 
recall that, for every vector field $K$ on an arbitrary Lorentzian 
manifold, the 3-form 
\begin{equation}\label{eqn-defomega}
\omega \, = \, \beta \, \wedge \, d \beta
\end{equation}
measures the {\em twist\/} (or {\em vorticity\/}) of $K's$ integral
curves, where the one-form $\beta$ is defined as
\begin{equation}\label{eqn-defbeta}
\beta \, = \, g(K, \, \cdot \, )  
\end{equation}
and $\wedge$ denotes the antisymmetrized tensor product. This observation 
is based on the fact (see e.g. Westenholz \cite{we}, pp. 212)
%Schutz \cite{sc}, p. 155 
that, on every simply connected region of $M$, $\beta$ admits an integrating 
factor,
\begin{equation}\label{eqn-int}
\beta = f \, dh
\end{equation}
with some scalar functions $f$ and $h$, if and only if $\omega$ vanishes.
If $K$ (and thus $\beta$) has no zeros, (\ref{eqn-int}) expresses the fact 
that the integral curves of $K$ are orthogonal to hypersurfaces $h = {\mathrm{const}}$. 
After these preparations we are now ready to prove the following proposition.

\begin{proposition}\label{prop-killing}
Let $(M,g)$ be a Lorentzian manifold and $K$ be a conformal
Killing vector field on $M$,
\begin{equation}\label{eqn-killing}
L_K g = \phi \, g
\end{equation}
where $L_K$ denotes the Lie derivative in the direction of $K$
and $\phi$ is some scalar function on $M$. Define $\beta$ and
$\omega$ according to $(\ref{eqn-defbeta})$ and $(\ref{eqn-defomega})$.
Let $I$ be an open interval and $\mu : I \longrightarrow M$ a lightlike 
geodesic such that 
\begin{equation}\label{eqn-twistcrit}
\omega \big( K_{\mu (s)}, {\dot{\mu}} (s) , \, \cdot \, \big) 
\, = \, 0
\end{equation}
and
\begin{equation}\label{eqn-transcrit}
g \big( K_{\mu (s)} , {\dot{\mu}} (s) \big) \neq 0
\end{equation}
for all $s \in I$. Define $\Sigma$ 
as the set of all points in $M$ that can be connected to $\mu$
by an integral curve of $K$. Then $\Sigma$ is a timelike photon 
$2$-surface.
\end{proposition}
\begin{proof}
Condition (\ref{eqn-transcrit}) makes sure that $K_{\mu (s)}$
and ${\dot{\mu}}(s)$ are linearly independent, so $\Sigma$ is a
2-dimensional submanifold of $M$. ($\Sigma$ might fail to be an 
{\em imbedded\/} submanifold because $\mu$ or an integral curve
of $K$ might be closed or almost closed.) Let $k$ denote the 
restriction of $K$ to $\Sigma$ and define a vector field $l^+$
on $\Sigma$ by
\begin{gather}
\label{eqn-lk}
[l^+,k] = 0  \, ,
\\
\label{eqn-lrestr}
l^+ _{\mu (s)} = {\dot{\mu}} (s) \qquad \text{for all} \quad s \in I \, .
\end{gather}
As the flow of a conformal Killing vector field maps lightlike
geodesics onto lightlike geodesics, $l^+$ is lightlike and geodesic
on all of $\Sigma$,
\begin{equation}\label{eqn-lcond}
g ( l^+, l^+ ) = 0 \qquad {\mathrm{and}} \qquad
\nabla _{l^+} l^+ \, = \, q \, l^+
\end{equation}
with some scalar function $q$ on $\Sigma$. Moreover, (\ref{eqn-killing})
and (\ref{eqn-lk}) imply that 
\begin{equation}\label{eqn-gkl}
k \, g(k, l^+ ) \, = \, \phi \, g (k, l^+ ) \, ,
\end{equation}
which, together with (\ref{eqn-transcrit}), makes sure that
\begin{equation}\label{eqn-gkln0}
g(k,l^+) \, \neq \, 0
\end{equation}
on all of $\Sigma$. With (\ref{eqn-gkln0}) being established,
the vector field
\begin{equation}\label{eqn-defl2}
l^- = \frac{g(k,k)}{g(k,l^+)^2} \, l^+ \, - \, 
\frac{2}{g(k,l^+)} \, k
\end{equation}
is well-defined on $\Sigma$. Clearly, $l^-$ is lightlike and
linearly independent of $l^+$. This implies that $\Sigma$
is timelike. What remains to be shown is that $l^-$ is 
geodesic. To prove this, we first evaluate our assumption
(\ref{eqn-twistcrit}). Together with (\ref{eqn-killing}) and 
(\ref{eqn-lk}), this assumption guarantees that 
\begin{equation}\label{eqn-omegakl}
\omega (k, l^+ , \, \cdot \, ) \, = \, 0
\end{equation}
on all of $\Sigma$. Hence, we have for all vector fields $Y$ 
along $\Sigma$ (which need not be tangent to $\Sigma$)
\begin{equation}\label{eqn-twistcrit2}
\begin{split}
0 \, = \, & g(k,k) \, d \beta (l^+,Y) \,  
+ g(k, l^+) \, d \beta (Y, k) \, + \, 
g(k,Y) \, d \beta (k, l^+)
\\
= \, & g(k,k) \, \Big( g \big(\nabla _{l^+} k , Y \big) 
- g \big( \nabla _Y k , l^+ \big) \Big) \, +
\, g(k,l^+) \, \Big( g \big(\nabla _Y k , k \big) \, - \, 
g \big( \nabla _k k , Y \big) \Big) 
\\
 & + \, g(k,Y) \, \Big( g \big(\nabla _k k , l^+ \big) 
- g \big( \nabla _{l^+} k , k \big) \Big)
\\
 = \, & g(k,k) \, \Big( 2 \, g \big(\nabla _{l^+} k , Y \big) 
 - \phi \, g ( Y, l^+ ) \Big) \, - 
 \, 2 \, g(k,l^+) \, g(\nabla _k k , Y) 
\\
 & + \, 2 \, g(k, Y) \, g( \nabla _k k , l^+ ) \, ,
\end{split}
\end{equation}
where, in the last step, (\ref{eqn-killing}) was used. As
(\ref{eqn-twistcrit2}) holds for all $Y$ along $\Sigma$,
we have found that
\begin{equation}\label{eqn-twistcrit3}
\begin{split}
0 \,  = \, & 2 \, g(k,k) \, \nabla _{l^+} k \, - \, 
\phi \, g(k,k) \, l^+  \, - \, 2 \, g(k, l^+) \, \nabla _k k 
\\
 + \, & 2 \, \phi \, g(k, l^+) \, k \, - \, 
 2 \, g( k, \nabla _k l^+ ) \, k
\end{split}
\end{equation}
along $\Sigma\,$; this equation will be used later . -- 
We now calculate $\nabla _{l^-} l^-$ from (\ref{eqn-defl2}). Using 
repeatedly (\ref{eqn-killing}), (\ref{eqn-lk}) and (\ref{eqn-lcond}) 
we find
\begin{equation}\label{eqn-dell2}
\begin{split}
\nabla _{l^-} l^- \,  = \, & \frac{g(k,k)}{g(k,l^+)^2} \left(
 \Big( l^+ \frac{g(k,k)}{g(k,l^+)} \Big) \, \frac{l^+}{g(k,l^+)} 
 \, - \, 2 \, \nabla _{l^+} \frac{k}{g(k,l^+)} \right) 
\\ 
 & - \; \frac{2}{g(k,l^+)} \, \left( \frac{g(k,k)}{g(k,l^+)} \,
 \nabla _k \frac{l^+}{g(k,l^+)} \, - \, 2 \, \nabla _k
 \frac{k}{g(k,l^+)} \right)
\\
  = \, & \left( l^+ \frac{g(k,k)}{g(k,l^+)} \, + \, 
 \frac{2 \, k \, g(k, l^+)}{g(k,l^+)} \right) \,  
 \frac{l^-}{g(k,l^+)} 
\\
 & + \, \frac{4}{g(k,l^+)^3} \Big( g(k, \nabla _{l^+} k ) \, k
 \, - \, g(k,k) \, \nabla _{l^+} k \, + 
 \, g(k,l^+ ) \, \nabla _k k \Big) \, .
\end{split}
\end{equation}
By (\ref{eqn-defl2}) and (\ref{eqn-twistcrit3}), this 
demonstrates that $\nabla _{l^-} l^-$ is a multiple of $l^-$.
\end{proof}

This proposition applies, in particular, to conformally static
spacetimes, i.e., to spacetimes admitting a timelike conformal
Killing vector field that is hypersurface-orthogonal. If $K$
is hypersurface-orthogonal, $\omega$ vanishes, so 
(\ref{eqn-twistcrit}) is automatically satisfied. If $K$ is
timelike, (\ref{eqn-transcrit}) is automatically satisfied.
Hence, in a conformally static spacetime Proposition
\ref{prop-killing} associates with {\em every\/} lightlike
geodesic a timelike photon 2-surface. In combination with
Theorem \ref{theo-main}, this gives the following observation
which is exemplified by the work of Abramowicz mentioned in
the introduction.  In a static or conformally static spacetime, 
one does not feel an inertial force directed sidewards if one 
moves along the spatial path of a light ray.

We emphasize that (\ref{eqn-twistcrit}) is automatically satisfied
whenever $K$ is hypersurface-orthogonal, independent of the
causal character of $K$. Note, however, that (\ref{eqn-transcrit})
gives a (mild) restriction on the allowed lightlike geodesics
$\mu$ if $K$ is not timelike. Taking this restriction into
account, we can apply Proposition \ref{prop-killing}, e.g., to 
the spacelike and hypersurface-orthogonal Killing vector field 
$\partial _{\varphi}$ in a static and axisymmetric spacetime.

If $K$ is not hypersurface-orthogonal, (\ref{eqn-twistcrit})
is a severe restriction on the allowed lightlike geodesics
$\mu$. In a stationary but non-static spacetime only those
lightlike geodesics $\mu$ give rise to a timelike photon 2-surface
for which ${\dot{\mu}} (s)$ points in the direction of the
local rotation axis of the timelike Killing vector field
for all parameter values $s$.

One might be tempted to
conjecture that {\em all\/} timelike photon 2-surfaces can 
be constructed with the help of conformal Killing vector fields
via Proposition \ref{prop-killing}, i.e., that timelike
photon 2-surfaces do not exist in spacetimes without conformal
Killing vector fields. We will now prove some results which demonstrate
that this conjecture is false. To that end we have to characterize
timelike photon surfaces in terms of their second fundamental form. 
Recall that the {\em second fundamental form\/} (or {\em shape tensor 
field\/}) $\Pi$ is well-defined for any nowhere lightlike submanifold 
$\Sigma$ of a semi-Riemannian manifold, in particular for a timelike 
submanifold of a Lorentzian manifold, by the equation
\begin{equation}\label{eqn-defPi}
\Pi (u, w) \, = \, P^{\perp} \big( \nabla _u w \big) \, ,
\end{equation}
where $P^{\perp}$ denotes the tensor field along $\Sigma$ which
is defined by the orthogonal projections (\ref{eqn-Pp}) and 
$u$ and $w$ are vector fields on $\Sigma$. As $[u,w]$ must be  
tangent to $\Sigma$, it is easy to verify the well-known fact that
$\Pi$ is a symmetric tensor field along $\Sigma$. For our 
purpose, the following notion is of particular interest, cf., e.g., 
O'Neill \cite{o}, p. 106.

\begin{definition}\label{def-umbilic}
A nowhere lightlike submanifold $\Sigma$ of a semi-Riemannian
manifold $(M,g)$ is called {\em totally umbilic\/} if there is a 
normal vector field $Z$ along $\Sigma$ such that 
\begin{equation}\label{eqn-umbilic}
\Pi (u, w) = g(u,w) \, Z
\end{equation}
for all vector fields $u,w$ on $\Sigma$. 
\end{definition}

A totally umbilic submanifold with $Z=0$ is called {\em totally 
geodesic\/}. It is easy to verify that the property of being
totally umbilic is conformally invariant whereas the property 
of being totally geodesic is not.  As a matter of fact, any totally  
umbilic submanifold can be locally converted into a totally geodesic
one by a suitable conformal transformation. Totally geodesic timelike  
2-surfaces have found a lot of interest since Vickers \cite{v} and others
demonstrated that the worldsheet of a self-gravitating cosmic string 
must be totally geodesic. Apparently there is not much interest
among relativists in totally umbilic timelike 2-surfaces up to now.

We will now prove that a 2-dimensional timelike submanifold of a Lorentzian 
manifold is a photon 2-surface if and only if it is totally umbilic. 
For timelike hypersurfaces, i.e., for the case dim$(M)=3$, the 
same result can be deduced from Claudel, Virbhadra and Ellis 
\cite{cve}, Theorem 2.2. 

\begin{proposition}\label{prop-umbilic}
Let $\Sigma$ be a $2$-dimensional timelike submanifold 
of a Lorentzian manifold $(M,g)$. Then $\Sigma$ is a photon 
$2$-surface if and only if $\Sigma$ is totally umbilic.
\end{proposition} 
\begin{proof}
Let $l^+$ and $l^-$ denote the two lightlike vector fields
on $\Sigma$, unique up to transformations of the form
(\ref{eqn-transl}), which are normalized according to 
(\ref{eqn-lnorm}). If $\Sigma$ is totally umbilic, (\ref{eqn-umbilic})
requires that $\Pi (l^{\pm},l^{\pm}) = 0$. By (\ref{eqn-defPi})
and (\ref{eqn-dell}), this implies that $l^+$ and $l^-$ are
geodesic, so $\Sigma$ is a photon 2-surface. -- Conversely,
assume that $\Sigma$ is a photon 2-surface. Consider any 
two vector fields $u =  \alpha \, l^+ \, + \beta \, l^-$
and $w = \gamma \, l^+ \, + \delta \, l^-$ on $\Sigma$,
where $\alpha, \beta , \gamma , \delta$ are scalar functions
on $\Sigma$. As $\nabla _{l^+} l^+$ and $\nabla _{l^-} l^-$
are tangent to $\Sigma$, the second fundamental form 
(\ref{eqn-defPi}) reduces to
\begin{equation}\label{eqn-PiZ}
\Pi (u,w) \, = \, 
\alpha \, \delta \, P^{\perp} \big( \nabla _{l^+} l^- \big) \, + 
\, \beta \, \gamma \, P^{\perp} \big( \nabla _{l^-} l^+ \big) \, .
\end{equation}
The fact that with $l^+$ and $l^-$ also the Lie bracket $[l^+,l^-]$ 
must be tangent to $\Sigma$ implies  
\begin{equation}\label{eqn-Zrep}
P^{\perp} \big( \nabla _{l^+} l^- \big) \, = 
\, P^{\perp} \big( \nabla _{l^-} l^+ \big) \, =: \, - \, 2 \, Z \, . 
\end{equation}
As $g(u,w) \, = \, - \, 2 \, \alpha \, \delta \, - \, 2 \,
\beta \, \gamma$, (\ref{eqn-PiZ}) now demonstrates that $\Pi$
has, indeed, the form (\ref{eqn-umbilic}).
\end{proof}

Based on this proposition we can now address the question of whether
a spacetime can be foliated into photon 2-surfaces. For Riemannian
manifolds (i.e., for positive definite metrics) it is known that 
even locally the existence of a foliation into totally umbilic submanifolds
is not guaranteed; criteria in terms of curvature conditions have been
given by Walschap \cite{wa}. Although, as far as we know, analogous results 
have not been worked out for Lorentzian or other indefinite signatures, 
the Riemannian results clearly indicate that only special classes of 
spacetimes admit a foliation into photon 2-surfaces. An interesting 
example is the class of twisted products. For local considerations,
this class can be defined in the following way. An $n$-dimensional 
pseudo-Riemannian manifold is locally a {\em twisted product\/} if it 
admits local coordinates $(u,v) = (u^1,\dots, u^m, v^1, \dots , v^{n-m})$ 
such that the metric takes the form
\begin{equation}\label{eqn-twist}
g \, = \, h_{ij}(u) \, du^i \, du^j \, + \,
\psi(u,v) \, k_{\mu \nu} (v) \, dv^{\mu} \, dv^{\nu}
\end{equation}
with summation over $i,j$ from 1 to $m$ and over $\mu, \nu$ from
1 to $n-m$. In the more special case that the ``twisting function'' $\psi$
is independent of $v$ one speaks of a {\em warped product}, cf. O'Neill
\cite{o}. It is an elementary exercise to verify that for any pseudo-Riemannian
metric of the form (\ref{eqn-twist}) the submanifolds $u = {\mathrm{const.}}$
are totally umbilic and the submanifolds $v = {\mathrm{const.}}$ are
totally geodesic. More generally, the following result is true. A 
pseudo-Riemannian manifold is locally a twisted product if and only if it 
locally admits two foliations ${\mathcal{F}}$ and ${\mathcal{G}}$ which are 
transverse and orthogonal to each other with all leaves of ${\mathcal{F}}$ 
totally geodesic and all leaves of ${\mathcal{G}}$ totally umbilic. For a 
proof we refer to Ponge and Reckziegel \cite{pr}; our statement is the local 
version of their Theorem 1. As a conformal factor has no influence upon the 
property of being totally umbilic, this observation together with Proposition 
\ref{prop-umbilic} gives us the following result. Every spacetime that is 
conformally related to a twisted product of a Riemannian factor and a 
2-dimensional Lorentzian factor is foliated into timelike photon 2-surfaces. 
As a twisted product need not admit any non-zero conformal Killing vector 
field, this demonstrates the existence of timelike photon 2-surfaces that 
cannot be constructed by the method of Proposition \ref{prop-killing}.

%%%%%%%%%%%%%%%%%%%%%%%%%%%%%%%%%%%%%%%%%%%%%%%%%%%%%%%%%%%%%%%%%%%%%%%%%%%%
%%%%%%%%%%%%%%%%%%%%%%%%%%%%%%%%%%%%%%%%%%%%%%%%%%%%%%%%%%%%%%%%%%%%%%%%%%%%
\section{Examples}
\label{sec-examples}

%----------------------------------------------------------------------

\subsection{Conformally flat spacetimes}
\label{subsec-flat}

As a photon 2-surface remains a photon 2-surface if the spacetime
metric is conformally transformed, the photon 2-surfaces in a 
conformally flat spacetime are locally the same as the photon
2-surfaces in Minkowski spacetime.

\begin{figure}
  \epsfig{file=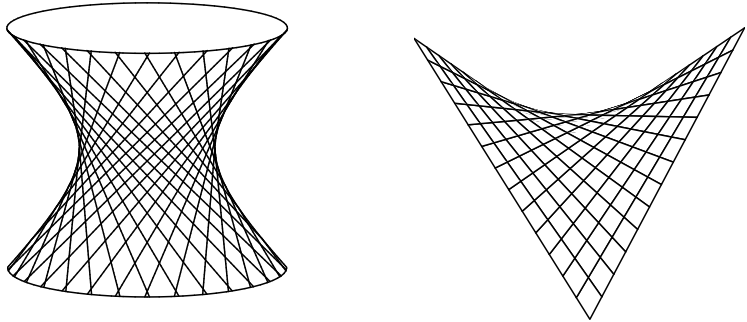,width=.95\linewidth} 
  \caption{In affine 3-space, the only surfaces other than planes 
   that admit two different rulings by straight lines are rotational 
   hyperboloids (left) and hyperbolic paraboloids (right). In
   3-dimensional Minkowski spacetime, the hyperbolic paraboloid 
   cannot be realized with lightlike rulings.}\label{fig-hyp}  
\end{figure}

To determine all timelike photon 2-surfaces in $n$-dimen\-sional 
Minkowski spacetime we begin with the case $n=3$. O'Neill 
\cite{o}, p. 117, proves that a connected, complete and nowhere
lightlike hypersurface in a vector space with non-degenerate 
scalar product of arbitrary signature is totally umbilic if and 
only if it is either a hyperplane or a hyperquadric. By Proposition
\ref{prop-umbilic}, this result implies that the connected and 
complete timelike photon 2-surfaces in 3-dimensional Minkowski 
spacetime are the timelike planes and the timelike quadrics. 
The latter are all timelike rotational hyperboloids, given in 
arbitrary Minkowski coordinates $(x,y,t)$ by an equation
of the form  
\begin{equation}\label{eqn-hyperbol}
(x-x_0)^2 \, + \, (y-y_0)^2 \, - \, (t-t_0)^2 \, =  \, r^2 \; ,
\end{equation}
where $x_0$, $y_0$, $t_0$ and $r>0$ are real numbers, see Figure \ref{fig-hyp}, 
left. A timelike plane can, of course, be interpreted as the worldsheet of a 
straight track in inertial motion. A timelike rotational hyperboloid can be interpreted
as the worldsheet of a circular track whose radius contracts and then re-expands 
in the course of time. Note that, owing to the invariance of (\ref{eqn-hyperbol})  
with respect to Lorentz boosts, this interpretation is independent of the choice
of a particular inertial system. It is obvious, intuitively, that someone moving 
along a track whose worldsheet is a plane does not feel centrifugal or Coriolis forces, 
and that the track appears straight to the eye and when probed with a gyroscope. Our 
results show the less obvious fact that the same is true for a track whose worldsheet 
is a rotational hyperboloid, and for no other tracks in 3-dimensional 
Minkowski spacetime.

By adding more spatial dimensions, a photon 2-surface in
3-dimensional Min\-kowski spacetime can, of course, also
be interpreted as a photon 2-surface in $n$-dimen\-sional
Minkowski spacetime for $n>3$. This construction gives, indeed, 
{\em all\/} photon 2-surfaces in $n$-dimensional Minkowski 
spacetime for arbitrarily large $n$. In order to prove this
one has to demonstrate that a totally umbilic timelike 2-surface
in $n$-dimensional Minkowski spacetime must be completely 
contained in a $3$-dimensional affine subspace. This
demonstration can be found in a paper by Hong \cite{ho}. 
Related results for pseudo-Euclidean spaces of arbitrary
signature are given by Ahn, Kim and Kim \cite{akk}.

Note that a timelike plane in Min\-kowski spacetime can be constructed 
by applying a timelike translation to a lightlike straight line.
Similarly, a timelike rotational hyperboloid can be constructed
by applying a spatial rotation to a lightlike straight line.
Hence, all photon 2-surfaces in conformally flat spacetimes can be 
constructed by the method of Proposition  \ref{prop-killing}. 

In addition, it is interesting to mention that a photon 2-surface
in $n$-dimen\-sional Minkowski spacetime is, in particular, a 
surface in $n$-dimensional affine space that admits two different
rulings by straight lines. It is well known that in 3-dimensional
affine space the only surfaces with this property are planes, rotational 
hyperboloids and hyperbolic paraboloids, see Figure \ref{fig-hyp}; for
a discussion of this classical result the reader may consult
Hilbert and Cohn-Vossen \cite{hc}. 
This observation gives an alternative way of determining all
timelike photon 2-surfaces in 3-dimensional Minkowski
spacetime (and thereupon, by the above argument, in $n$-dimensional
Minkowski spacetime): One just has to single out those doubly ruled 
surfaces for which both rulings are realized by lightlike lines. It
is an elementary exercise to verify that the only surviving
surfaces are timelike planes and timelike rotational hyperboloids.

%------------------------------------------------------------------

\subsection{Schwarzschild and Reissner-Nordstr{\"o}m spacetimes}
\label{subsec-schwarzschild}

In this subsection we consider $(2+1)$-dimensional spacetimes
with metrics of the form
\begin{equation}\label{eqn-spherical}
g \; = \; - \, e^{2 \Phi (r)} \, dt^2 \, + \,
e^{-2 \Phi (r)} \, dr^2 \, + \, 
r^2 \, d{\varphi}^2 \, .
\end{equation}
This includes, as particular examples, the restriction to the
equatorial plane of the Schwarzschild spacetime,
\begin{equation}\label{eqn-schwarzschild}
e^{2 \Phi (r)} \, = \, 1 \, - \, \frac{2M}{r} \, ,
\end{equation}
and of the Reissner-Nordstr{\"o}m spacetime,
\begin{equation}\label{eqn-reissner}
e^{2 \Phi (r)} \, = \, 1 \, - \, \frac{2M}{r} \, + \, \frac{Q^2}{r^2} \, .
\end{equation}
As both $\partial _t$ and $\partial _{\varphi}$ are hypersurface-orthogonal
Killing vector fields, Proposition \ref{prop-killing} allows to construct 
$\partial _t$-invariant photon 2-surfaces and $\partial _{\varphi}$-invariant 
photon 2-surfaces. The former are associated with static tracks, the latter 
with circular tracks that contract or expand in the course of time. Here it 
is our goal to determine the $\partial _{\varphi}$-invariant photon 2-surfaces.
Instead of using Proposition \ref{prop-killing}, we find it more convenient 
to construct these photon 2-surfaces with the help of Proposition 
\ref{prop-frobenius}.

As an ansatz, we consider two vector fields
\begin{equation}\label{eqn-Lplumi}
L^{\pm} \, = \, A(r) \, \Big( \, \partial _t \, + \, 
X(r) \, \partial _r \, \Big) \, \pm \, \frac{1}{r} \, 
\partial _{\varphi} \, ,
\end{equation}
where $A$ and $X$ are functions of $r$ to be specified
later. The Lie bracket of these vector fields satisfies
\begin{equation}\label{eqn-lie}
[L^+,L^-] \, = \, \frac{A(r) \, X(r)}{r} \, 
\big( \, L^+ \, - \, L^- \, \big) \, ,
\end{equation}
which demonstrates that $L^+$ and $L^-$ are surface-forming for 
any choice of $A$ and $X$. The additional condition
\begin{equation}\label{eqn-norm}
X(r)^2 \, = \, e^{2 \Phi(r)} \, 
\big( \, e^{2\Phi (r)} \, - \, A(r)^{-2} \, \big)
\end{equation}
makes sure that $L^+$ and $L^-$ are lightlike. This implies
that the 2-dimensional integral manifolds generated by $L^+$
and $L^-$ are timelike. As each of these integral manifolds
is obviously invariant under the flow of $\partial _{\varphi}$,
it can be interpreted as the worldsheet of a circular track
whose radius changes with time. We now add the condition that
$L^+$ and $L^-$ be geodesic,
\begin{equation}\label{eqn-geo}
\nabla _{L^{\pm}} L^{\pm} \, = \, f_{\pm} \, L^{\pm} 
\end{equation}
with some functions $f_+$ and $f_-$.
This is true if and only if
\begin{equation}\label{eqn-geocoo}
g \big( \nabla _{L^{\pm}} L^{\pm} , \partial _i \big)
\,  = \, f_{\pm} \, g ( L^{\pm} , \partial _i )
\end{equation}
for $i= \varphi , t , r \, $. As $L^{\pm}$ is lightlike,
we have 
\begin{equation}\label{eqn-geoi}
g \big( \nabla _{L^{\pm}} L^{\pm} , \partial _i \big)
\,  = \, L^{\pm} \, g(L^{\pm},\partial _i ) \, - \,
g(L^{\pm},[L^{\pm}, \partial _i ]) \, .
\end{equation}
With (\ref{eqn-geoi}) it is easy to evaluate (\ref{eqn-geocoo}). 
For $i = \varphi$ and $i=t$ we find respectively
\begin{gather}
\label{eqn-phi}
A(r) \, X(r) \, = \, f_{\pm} \, r
\\
\label{eqn-t}
X(r) \, A(r) \, A'(r) \, - \, 2 \, X(r) \, A(r)^2 \, \Phi '(r) \, = \,
f_{\pm} \, A(r) \,  .
\end{gather}
The equation for $i=r$ turns out to be a consequence of (\ref{eqn-norm}), 
(\ref{eqn-phi}) and (\ref{eqn-t}), so we do not have to consider it. 
Upon eliminating $f_{\pm}$ from (\ref{eqn-phi}) and (\ref{eqn-t}) 
we get an ordinary first-order differential equation for $A$
which can be solved easily, yielding
\begin{equation}\label{eqn-A}
A(r) \; = \; \frac{r}{R} \; e^{-2\Phi (r)} \, ,
\end{equation}
where, without loss of generality, the constant of integration $R$ 
can be assumed positive. For any choice of $R \, (\, > 0\, )$, inserting
(\ref{eqn-A}) into (\ref{eqn-norm}) gives us a positive and a negative 
solution $X$ on that part of the spacetime where 
\begin{equation}\label{eqn-ineq}
r^2 \, e^{-2 \, \Phi (r)} \, > \, R^2 \, .
\end{equation}
If (\ref{eqn-ineq}) holds on the whole spacetime under
consideration, this construction gives us two families of photon 
2-surfaces, one corresponding to the positive solution for $X$ 
and one corresponding to the negative solution for $X$, such that
the members of either family foliate the spacetime. 
If (\ref{eqn-ineq}) is true only on a proper subset $U$ of the 
spacetime, then it is not difficult to verify that all members 
of both families either meet the boundary $\partial U$ tangentially 
or asymptotically approach the boundary $\partial U$. In the first 
case each member of the first family is to be glued together with a 
member of the second family at the boundary. So there is actually 
only one family of photon 2-surfaces which covers $U$ twice. 

\begin{figure}
  \epsfig{file=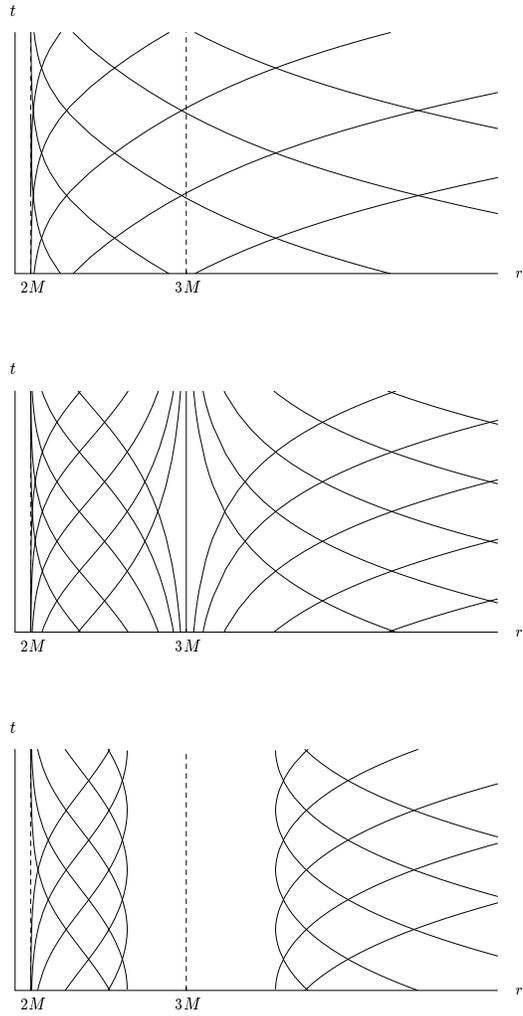,width=.95\linewidth} 
  \caption{$\partial _{\varphi}$-invariant timelike photon 2-surfaces 
  in Schwarzschild spacetime}\label{fig-psurf}  
\end{figure}

These results are best illustrated with the exterior
Schwarzschild spacetime. To that end we specify $\Phi$ 
according to (\ref{eqn-schwarzschild}) and we restrict to the
region $r>2M$. For $R<\sqrt{27} \, M$, (\ref{eqn-ineq})
is satisfied for all $r>2M$. Hence, $R$ is associated with
two different families of photon 2-surfaces. Each of these photon 
2-surfaces is generated by lightlike geodesics that extend from the 
horizon at $r=2M$ to infinity, see Figure \ref{fig-psurf}, top. For  
$R= \sqrt{27} \, M$, (\ref{eqn-ineq}) is true everywhere on $r>2M$ 
with the exception of the value $r=3M$ where the left-hand side of 
(\ref{eqn-ineq}) is equal to the right-hand side. It is still true 
that $R$ is associated with two different families of photon 2-surfaces;
however, the surface $r=3M$ belongs to both families. All other members 
of both families asymptotically approach this particular photon 2-surface 
at $r=3M$ either for $t \to \infty$ or for $t \to - \infty$, see 
Figure \ref{fig-psurf}, middle. For $R>\sqrt{27} \, M$, (\ref{eqn-ineq}) 
is true for all $r>2M$ with the exception of some interval around $r = 3M$. 
There is only one family of photon 2-surfaces associated with $R$ which 
covers the allowed region twice. Each member of this family is ruled by 
lightlike geodesics which either come from infinity, reach a minimal 
radius and go out to infinity, or come from the horizon, reach a maximal 
radius and go down to the horizon, see Figure \ref{fig-psurf}, bottom. 

%--------------------------------------------------------------------

\subsection{G\"odel spacetime}
\label{subsec-goedel}

The G\"odel metric
\begin{equation}\label{eqn-goedel}
g \, = \, - \, dt^2 \, + dx^2 \, - \, \tfrac{1}{2} \, e^{2 \sqrt{2} \,
\omega _o x} dy^2 \, + \, dz^2 \, - \, 2 \, e^{\sqrt{2} \, \omega _o x} 
dt \, dy
\end{equation}
is a rotating dust solution of Einstein's field equation with cosmological 
constant, cf., e.g., Hawking and Ellis \cite{he}, p. 168.
A foliation of the G\"odel spacetime into timelike photon 2-surfaces can be
easily found with the help of Proposition \ref{prop-frobenius}. To that
end, we observe the obvious fact that the vector fields
\begin{equation}\label{eqn-Lgoedel}
L^{\pm} \, = \, \partial _t \, \pm \, \partial _z
\end{equation}
are lightlike and have vanishing Lie bracket, $[L^+,L^-] = 0$.
Moreover, for all basis vector fields $\partial _i$ with $i =
t,z,x,y$ we read from (\ref{eqn-goedel})
\begin{equation}\label{eqn-goedelgeo}
g \big( \nabla _{L^{\pm}} L^{\pm} , \, \partial _i \, \big) 
\, = \, L^{\pm} g(L^{\pm},\partial _i ) \, - \,
\tfrac{1}{2} \, \partial _i \big( g(L^{\pm},L^{\pm}) \big)\, = \, 0 \, ,
\end{equation}
so $\nabla _{L^ {\pm}}L^{\pm} = 0$. By Proposition \ref{prop-frobenius},
the surfaces $x = {\mathrm{const.}}, y = {\mathrm{const.}}$ are
timelike photon 2-surfaces. At the same time, these timelike
photon 2-surfaces exemplify Proposition \ref{prop-killing},
where $K$ is to be identified with the Killing vector field
$\partial _t$. To apply Proposition \ref{prop-killing} one just
has to verify from (\ref{eqn-goedel}) that for   $\beta = 
g (\partial _t, \, \cdot \, )$ the twist 3-form $\omega = \beta 
\wedge d \beta$ satisfies $\omega (\partial _t, \partial _z , \, \cdot \, ) 
=0\,$; thus, (\ref{eqn-twistcrit}) is indeed true for $\dot{\mu}$ parallel
to $L^+$ or $L^-$.

It is interesting to note that this foliation into timelike photon 
2-surfaces of the G\"odel spacetime is {\em not\/} associated with a 
twisted-product structure. This is obvious from the fact that, because 
of the $dy \, dt$-term in the metric (\ref{eqn-goedel}), the 2-spaces 
orthogonal to the surfaces $x = {\mathrm{const.}}, y = {\mathrm{const.}}$ 
are not integrable.

The surfaces $x = {\mathrm{const.}}, y = {\mathrm{const.}}$ are not 
the only timelike photon 2-surfaces in the G\"odel cosmos. Other 
examples can be found, e.g., by applying Proposition \ref{prop-killing}
to the hypersurface-orthogonal spacelike Killing vector field  
$K = \partial _z$ and to any lightlike geodesic $\mu$ that is not
perpendicular to $\partial _z$.

%%%%%%%%%%%%%%%%%%%%%%%%%%%%%%%%%%%%%%%%%%%%%%%%%%%%%%%%%%%%%%%%%%%%%%%%%%%%
%%%%%%%%%%%%%%%%%%%%%%%%%%%%%%%%%%%%%%%%%%%%%%%%%%%%%%%%%%%%%%%%%%%%%%%%%%%%

\end{document}